\documentclass[amsmath,amssymb,aps,pra,reprint,groupedaddress,showpacs]{revtex4-1}

\usepackage{multirow}
\usepackage{verbatim}
\usepackage{color,graphicx}

\newcommand{\Proof}{\noindent\textbf{Proof.}\quad}
\newcommand{\qed}{\hfill$\Box$}
\newcommand{\ket}[1]{\left\vert #1 \right\rangle}
\newcommand{\bra}[1]{\left\langle #1 \right\vert}

\newtheorem{theorem}{Theorem}
\newtheorem{lemma}[theorem]{Lemma}

\newtheorem{proposition}[theorem]{Proposition}

\begin{document}

\title{Using arbitrary parity-check matrices for quantum error correction assisted by less noisy qubits}

\author{Yuichiro Fujiwara}
\email[]{yuichiro.fujiwara@caltech.edu}
\affiliation{Division of Physics, Mathematics and Astronomy, California Institute of Technology, MC 253-37, Pasadena, California 91125, USA}

\date{\today}

\begin{abstract}
Recently a framework for assisted quantum error correction was proposed in which a specific type of error is allowed to occur on auxiliary qubits,
which is in contrast to standard entanglement assistance that requires noiseless auxiliary qubits.
However, while the framework maintains the ability to import any binary or quaternary linear code without sacrificing active error correction power,
it requires the code designer to turn a parity-check matrix of the underlying classical code into an equivalent one in standard form.
This means that classical coding theoretic techniques that require parity-check matrices to be in specific form may not fully be exploitable.
Another issue of the recently proposed scheme is that the error correction capabilities for bit errors and phase errors are generally equal,
which is not ideal for asymmetric error models.
This paper addresses these two problems.
We generalize the framework in such a way that any parity-check matrix of any binary or quaternary linear code can be exploited.
Our generalization also allows for importing a pair of distinct linear codes so that error correction capabilities become suitably asymmetric.
\end{abstract}

\pacs{03.67.Pp, 03.67.Hk}

\maketitle

\section{Introduction}\label{intro}
Quantum error correction is critically important in making practical quantum information processing more realistic
because qubits, which are the carrier of quantum information, are inevitably subject to undesirable noise.
For this reason, a multitude of aspects of quantum error correction have extensively been studied \cite{Lidar:2013}.

In the classical domain, the theory of error correction has produced a wide range of promising schemes that have successfully been implemented in real world applications.
Those schemes that detect and correct unwanted changes in information are called \textit{error-correcting codes}.
Analogously, \textit{quantum error-correcting codes} are schemes whose aim is to suppress noise on qubits and correct unintended changes in quantum states.

Perhaps not surprisingly, designing a desirable error-correcting code appears more challenging in the quantum domain than in the classical domain.
For instance, the \textit{stabilizer formalism} of quantum error correction is a classic general framework
that has been investigated from various angles \cite{Gottesman:1997}.
Quantum error-correcting codes based on this formalism are called \textit{stabilizer codes}
and may be regarded as a quantum analogue of linear or additive codes in classical coding theory.
In view of the similarity, one might hope to extend sophisticated error correction techniques from classical coding theory to the quantum case.
This is partially possible through the stabilizer formalism.
However, the quantum analogues of advanced and practical error-correcting codes are often hampered by peculiarities of the framework.

Since the appearance of the stabilizer formalism,
various other frameworks for quantum error correction have been developed as well,
including the \textit{entanglement-assisted stabilizer formalism} \cite{Brun:2006,Brun:2014}, \textit{operator quantum error correction} \cite{Kribs:2005,Kribs:2006},
\textit{codeword stabilized formalism} \cite{Cross:2009}, and their unifications \cite{Hsieh:2007,Shin:2011,Shin:2013}, to name a few.
Among the recent breakthroughs, entanglement assistance is an interesting development in the study of direct quantum analogues of classical error-correcting codes
because it shows that entanglement, which is unique to the quantum domain, allows for
directly importing any binary or quaternary linear codes for active quantum error correction.
Hence, in theory, we may exploit many of the state-of-the-art techniques in classical coding theory in a simple manner in the quantum domain as well.

The major disadvantage of entanglement assistance, however, is that it generally assumes that auxiliary qubits are completely free from errors.
While not all excellent classical error-correcting codes require too many auxiliary qubits,
it is a challenging task to realize such noiseless qubits, however few they may be \cite{Hsieh:2011,Fujiwara:2010e}.
Therefore, it is important to study how the presence of noise on auxiliary qubits affects assisted quantum error correction \cite{Lai:2012}.

Recently the author proposed a different framework of assisted quantum error correction that addresses this problem,
where any binary or quaternary linear code can be turned into a quantum error-correcting code
without sacrificing active error correction power while allowing a specific type of error to occur on auxiliary qubits \cite{Fujiwara:2013c}.
In a sense, it is a form of quantum error correction assisted by ``less noisy'' qubits rather than noiseless ones.
For instance, because phase damping is expected to be much harder to suppress on hardware in typical quantum devices,
it is natural to assume that phase errors would still occur on auxiliary qubits
even if they are engineered more reliably or protected more tightly than other qubits \cite{Ioffe:2007,Sarvepalli:2009}.
In the recently proposed framework, one may allow this dominant type of noise to disturb auxiliary qubits.
Hence, the burden of making auxiliary qubits error-free is alleviated to a more feasible task of eliminating the kinds of error that are easier to suppress.

However, it is not a complete replacement of entanglement assistance.
One drawback of the scheme assisted by less noisy qubits is that
it requires the code designer to turn a parity-check matrix of the underlying classical code into an equivalent one in standard form
before creating the corresponding quantum error-correcting code.
Because one linear code admits multiple equivalent parity-check matrices in general,
this means that assistance by less noisy qubits may not be able to effectively exploit classical coding theoretic techniques
that require parity-check matrices to be of some specific form.
For instance, the \textit{sum-product algorithm} is a very efficient decoding method
whose error correction performance can come close to that of the much more computationally demanding maximum likelihood decoding \cite{MacKay:2003}.
However, it is effective only if parity-check matrices are of special sparse form.
Hence, while the assisted scheme can use any linear code in principle,
this specific decoding algorithm may not be effective if its parity-check matrix in standard form happens to be unsuitable.

Another weakness of the framework is that the error correction capabilities are generally symmetric.
The assisted quantum error correction exploits a common technique called \textit{discretization} of quantum noise,
so that the decoder only needs to be able to correct bit errors and phase errors.
In this setting, symmetric quantum error-correcting codes have equal tolerance against each type of error on noisy qubits.
However, because it is likely that phase errors occur on qubits more frequently than bit errors,
the decoder may have to waste error correction power that is too strong for less likely errors under a realistic noise model
or otherwise be overwhelmed by a more frequent kind of error.

The purpose of this paper is to solve these problems by generalizing the framework of quantum error correction assisted by less noisy qubits.
We give a generalization that makes it possible to
import any parity-check matrix of any binary or quaternary linear code as in entanglement assistance
while keeping the same active quantum error correction power and the key feature that one type of error is allowed to occur on auxiliary qubits.
The generalized framework also allows for importing a pair of distinct linear codes with different levels of error correction power,
leading to naturally asymmetric error correction capabilities accordingly.

\section{Generalized assistance by less noisy qubits}
In this section, we present the details of our generalized version of assisted quantum error correction.
This section is divided into two subsections.
Section \ref{preliminaries} gives a brief review of classical coding theory we use.
Section \ref{mainresults} presents our generalized quantum error correction scheme assisted by less noisy qubits.
Throughout this section we use basic facts in classical coding theory and quantum error correction.
For more thorough treatments of the fundamentals of classical coding theory than the brief review that follows,
the reader is referred to \cite{MacWilliams:1977,Huffman:2003b}.
For the basic facts and notions in quantum error correction and quantum information in general, we refer the reader to \cite{Lidar:2013,Nielsen:2000}.

\subsection{Classical error-correcting codes}\label{preliminaries}
A \textit{linear} $[n,k,d]_q$ {code} of \textit{length} $n$, \textit{dimension} $k$, and \textit{minimum distance} $d$
is a $k$-dimensional subspace $\mathcal{C}$ of the $n$-dimensional vector space $\mathbb{F}_q^n$ over the finite field $\mathbb{F}_q$ of order $q$
such that $\min\left\{\operatorname{wt}(\boldsymbol{c}) \mid \boldsymbol{c} \in \mathcal{C}\setminus\{\boldsymbol{0}\}\right\} = d$,
where $\operatorname{wt}(\boldsymbol{c})$ is the number of nonzero entries of $\boldsymbol{c}$.
The vectors in $\mathcal{C}$ are the \textit{codewords}.
We use the finite field $\mathbb{F}_4 = \{0,1,\omega,\omega^2=\omega+1\}$ of order four and its prime subfield $\mathbb{F}_2 = \{0,1\}$.
In the binary case, we omit the subscript in the parameter notation and write $[n,k,d]$
to describe the parameters of a linear $[n,k,d]_2$ code.

A linear $[n,k,d]_q$ code can also be defined as the kernel $\left\{\boldsymbol{c}\in\mathbb{F}_q^n \mid H\boldsymbol{c}^T = 0\right\}$
of some $(n-k) \times n$ matrix $H$ of full rank over $\mathbb{F}_q$, called a \textit{parity-check matrix},
because the code is a $k$-dimensional subspace of the $n$-dimensional vector space.
A simple linear algebraic observation shows that a linear code of minimum distance $d$ can identify any combination of errors
that occurred on at most $\left\lfloor\frac{d-1}{2}\right\rfloor$ positions in a codeword \cite{MacWilliams:1977}.
This fact is closely related to the fundamental decoding method for linear codes, namely \textit{syndrome decoding}.
We briefly review the error correction mechanism
in a manner relevant to our generalized scheme for quantum error correction.

The \textit{trace function} $\operatorname{Tr}$ from $\mathbb{F}_4$ onto $\mathbb{F}_2$ is defined to be $\operatorname{Tr}(a) = a + a^2$ for $a \in \mathbb{F}_4$.
The \textit{trace} $\operatorname{Tr}(\boldsymbol{a})$ of an $n$-dimensional vector $\boldsymbol{a} = (a_0, \dots, a_{n-1}) \in \mathbb{F}_4^n$
is the $n$-dimensional vector $\operatorname{Tr}(\boldsymbol{a}) = \left(\operatorname{Tr}(a_0),\dots, \operatorname{Tr}(a_{n-1})\right) \in \mathbb{F}_2^n$.
Any vector $\boldsymbol{a} \in \mathbb{F}_4^n$ can be expressed by using $\operatorname{Tr}$ as
$\boldsymbol{a} = \omega^2\operatorname{Tr}(\boldsymbol{a}) + \operatorname{Tr}(\omega\boldsymbol{a})$.
The trace function is $\mathbb{F}_4$-additive and $\mathbb{F}_2$-linear,
which means that for any $x, y \in \mathbb{F}_2$ and any $\boldsymbol{a}, \boldsymbol{b} \in \mathbb{F}_4^n$, we have
$\operatorname{Tr}(x\boldsymbol{a}+y\boldsymbol{b}) = x\operatorname{Tr}(\boldsymbol{a})+y\operatorname{Tr}(\boldsymbol{b})$.
We define the trace of a column vector in the same manner,
so that $\operatorname{Tr}\mkern-\medmuskip\left(\boldsymbol{a}^T\right) = \left(\operatorname{Tr}(\boldsymbol{a})\right)^T$.
The $\mathbb{F}_4$-additivity and $\mathbb{F}_2$-linearity of $\operatorname{Tr}$
implies that for any $\boldsymbol{a} \in \mathbb{F}_4^n$ and any binary matrix $A$ with $n$ columns, we have
$\textup{Tr}\mkern-\medmuskip\left(A\boldsymbol{a}^T\right) = A\left(\operatorname{Tr}(\boldsymbol{a})\right)^T$.

An \textit{additive} $(n, 2^{w},d)_4$ \textit{code} of \textit{length} $n$, $\mathbb{F}_2$-\textit{dimension} $w$, and \textit{minimum distance} $d$
over $\mathbb{F}_4$ is an $\mathbb{F}_2$-linear subset $\mathcal{C} \subseteq \mathbb{F}_4^n$
of the $n$-dimensional vector space over $\mathbb{F}_4$, where $\vert\mathcal{C}\vert = 2^w$
and $\min\left\{\operatorname{wt}(\boldsymbol{c}) \mid \boldsymbol{c} \in \mathcal{C}\setminus\{\boldsymbol{0}\}\right\} = d$.
Simply put, an additive $(n, 2^{w},d)_4$ code $\mathcal{C}$ over $\mathbb{F}_4$ is a set of vectors in $\mathbb{F}_4^{n}$ that is closed under addition,
of cardinality $2^w$, and of minimum distance $d$ as an error-correcting code.
Each element $\boldsymbol{c} \in \mathcal{C}$ is a \textit{codeword} of $\mathcal{C}$.

Linear codes of parameters $[n,k,d]_4$ and $[n,k,d]$ are additive codes of parameters $(n,4^{k},d)_4$ and $(n,2^k,d)_4$ respectively
while the converse may not be true for quaternary linear codes.
Similar to the fact that a linear code can be seen as the kernel of a parity-check matrix,
an additive $(n, 2^{w},d)_4$ code $\mathcal{C}$ can be specified by some $(2n-w) \times n$ matrix $H_Q$ over $\mathbb{F}_4$ as
\[\mathcal{C} = \left\{\boldsymbol{c}\in\mathbb{F}_4^n \mid \operatorname{Tr}\mkern-\medmuskip\left(H_Q\boldsymbol{c}^T\right) = 0\right\}.\]
The matrix $H_Q$ is called a \textit{trace parity-check matrix} of $\mathcal{C}$.
For an $n$-dimensional vector $\boldsymbol{a} \in \mathbb{F}_4^n$,
the column vector $\textup{Tr}\mkern-\medmuskip\left(H_Q\boldsymbol{a}^T\right)$ is the \textit{trace syndrome} of $\boldsymbol{a}$.
The following proposition shows that a trace parity-check matrix can be used to perform syndrome decoding.
\begin{proposition}\label{syndromedecoding}
Let $\mathcal{C}$ be an additive $(n,2^{w},d)_4$ code over $\mathbb{F}_4$ and $H_Q$ its trace parity-check matrix.
For any pair $\boldsymbol{e}, \boldsymbol{e}' \in \mathbb{F}_4^n$ of distinct $n$-dimensional vectors such that
$\operatorname{wt}(\boldsymbol{e}), \operatorname{wt}(\boldsymbol{e}') \leq \left\lfloor\frac{d-1}{2}\right\rfloor$,
their trace syndromes are distinct, that is, $\textup{Tr}\mkern-\medmuskip\left(H_Q\boldsymbol{e}^T\right) \not= \textup{Tr}\mkern-\medmuskip\left(H_Q{\boldsymbol{e}'}^T\right)$.
\end{proposition}
\Proof Suppose to the contrary that $\textup{Tr}(H_Q\boldsymbol{e}) = \textup{Tr}(H_Q\boldsymbol{e}')$
for some pair $\boldsymbol{e}, \boldsymbol{e}' \in \mathbb{F}_4^n$ of distinct $n$-dimensional vectors such that
$\operatorname{wt}(\boldsymbol{e}), \operatorname{wt}(\boldsymbol{e}') \leq \left\lfloor\frac{d-1}{2}\right\rfloor$.
Then because $\textup{Tr}$ is $\mathbb{F}_4$-additive,
we have
\begin{eqnarray*}
0 &=& \textup{Tr}\mkern-\medmuskip\left(H_Q\boldsymbol{e}^T\right) + \textup{Tr}\mkern-\medmuskip\left(H_Q{\boldsymbol{e}'}^T\right)\\
&=& \textup{Tr}\mkern-\medmuskip\left(H_Q(\boldsymbol{e}+\boldsymbol{e}')^T\right),
\end{eqnarray*}
which implies that $\boldsymbol{e}+\boldsymbol{e}' \in \mathcal{C}$.
However, because
\begin{eqnarray*}
\operatorname{wt}(\boldsymbol{e}+\boldsymbol{e}') &\leq&
\operatorname{wt}(\boldsymbol{e}) + \operatorname{wt}(\boldsymbol{e}')\\
&\leq& 2\left\lfloor\frac{d-1}{2}\right\rfloor\\
&\leq& d-1,
\end{eqnarray*}
this is a contradiction.
\qed

The importance of the above proposition lies in the fact that it is enough to compute the trace syndrome to be able to identify errors.
For instance, if a codeword $\boldsymbol{c}$ is altered to a different vector $\boldsymbol{c}'$,
we would like to identify the difference $\boldsymbol{e} = \boldsymbol{c} - \boldsymbol{c}'$.
Since $H_Q\boldsymbol{c}^T = 0$ and $-1 = 1$ in $\mathbb{F}_4$, we have $H_Q{\boldsymbol{c}'}^T = H_Q\boldsymbol{e}^T$.
Therefore, any discripancy $\boldsymbol{e}$ between $\boldsymbol{c}$ and $\boldsymbol{c}'$
can be identified as long as the number of errors, which is $\operatorname{wt}(\boldsymbol{e})$, does not exceed $\left\lfloor\frac{d-1}{2}\right\rfloor$.
Note that for a linear code, the same argument can be carried out by using its parity-check matrix $H$
and the \textit{syndrome} $H{\boldsymbol{c}'}^T$ rather than their trace variants.
We exploit the argument involving the trace function to import quaternary codes
while we let syndromes by parity-check matrices play this role in the binary case.

The trace syndrome can be computed through binary matrices.
Given a trace parity-check matrix $H_Q$ of an additive code of length $n$ and $\mathbb{F}_2$-dimension $w$,
there exists a unique decomposition $H_Q =  H_Z + \omega H_X$ into a pair $H_Z$, $H_X$ of $(2n-w)\times n$ matrices over $\mathbb{F}_2$.
We call $H_Z$ and $H_X$ the $Z$-\textit{matrix} and $X$-\textit{matrix} of $H_Q$ respectively.
\begin{proposition}\label{tracesyndromeexpression}
Let $H_Z$ and $H_X$ be the $Z$-matrix and $X$-matrix of a trace parity-check matrix $H_Q$ of an additive code $\mathcal{C}$ of length $n$ over $\mathbb{F}_4$.
For any $\boldsymbol{a} \in \mathbb{F}_4^n$, the trace syndrome can be expressed as
$\operatorname{Tr}\mkern-\medmuskip\left(H_Q\boldsymbol{a}^T\right) = H_Z\textup{Tr}\mkern-\medmuskip\left(\boldsymbol{a}^T\right)
+ H_X\textup{Tr}\mkern-\medmuskip\left(\omega\boldsymbol{a}^T\right)$.
\end{proposition}
\Proof Recall that any vector $\boldsymbol{a}$ over $\mathbb{F}_4$ can be expressed as
$\boldsymbol{a} = \omega^2\textup{Tr}(\boldsymbol{a}) +  \textup{Tr}(\omega\boldsymbol{a})$.
Because the trace function $\textup{Tr}$ is $\mathbb{F}_4$-additive and $\mathbb{F}_2$-linear, we have
\begin{align*}
\textup{Tr}\mkern-\medmuskip\left(H_Q\boldsymbol{a}^T\right)
&= \textup{Tr}\mkern-\medmuskip\left((H_Z+\omega H_X)(\omega^2\textup{Tr}(\boldsymbol{a})+\textup{Tr}(\omega\boldsymbol{a}))^T\right)\\
&= H_Z\textup{Tr}\mkern-\medmuskip\left(\omega^2\textup{Tr}\mkern-\medmuskip\left(\boldsymbol{a}^T\right)
+\textup{Tr}\mkern-\medmuskip\left(\omega\boldsymbol{a}^T\right)\right)\\
&\quad+ H_X\textup{Tr}\mkern-\medmuskip\left(\textup{Tr}\mkern-\medmuskip\left(\boldsymbol{a}^T\right)
+\omega\textup{Tr}\mkern-\medmuskip\left(\omega\boldsymbol{a}^T\right)\right)\\
&= H_Z\textup{Tr}\mkern-\medmuskip\left(\boldsymbol{a}^T\right) + H_X\textup{Tr}\mkern-\medmuskip\left(\omega\boldsymbol{a}^T\right)
\end{align*}
as desired.
\qed

\subsection{Assisted quantum error correction}\label{mainresults}
Now we describe how to import linear codes for quantum error correction.
Throughout this subsection, we assume that the type of noise allowed to affect auxiliary qubits is phase damping.
As in the original framework given in \cite{Fujiwara:2013c}, this can be modified so that the auxiliary qubits may only suffer from bit errors in a straightforward manner.

We first consider the case when the underlying classical code is quaternary.
We improve the original proof of the following theorem.
\begin{theorem}[\cite{Fujiwara:2013c}]\label{maintheorem}
If there exists a linear $[n,k,d]_4$ code over $\mathbb{F}_4$,
then there exit unitary operations that encode $k$ logical qubits into $2n-k$ physical qubits and correct up to $\left\lfloor\frac{d-1}{2}\right\rfloor$ errors
under the assumption that a fixed set of $2(n-k)$ physical qubits may experience phase errors but no bit errors.
\end{theorem}

The encoding and decoding operators used in the original proof of the above theorem require a parity-check matrix
that has a set of $n-k$ columns forming the $(n-k)\times(n-k)$ identity matrix.
While it is not difficult to see that all linear codes admit such parity-check matrices,
this requirement can make sophisticated classical coding theoretic techniques for efficient decoding less effective \cite{Fujiwara:2014}.
We prove that this condition on the form of parity-check matrices can be removed entirely.

Let $H$ be an $(n-k)\times n$ parity-check matrix of an $[n,k,d]_4$ linear code $\mathcal{C}$ over $\mathbb{F}_4$.
Because $H$ is full rank over $\mathbb{F}_4$, there exists $n-k$ linearly independent columns in $H$.
Without loss of generality, we assume that the first $n-k$ columns are linearly independent.
The following matrix
\begin{align}\label{tracepcm}
H_Q = \left[\begin{array}{c} H \\ \omega H\end{array}\right]
\end{align}
forms a trace parity-check matrix of $\mathcal{C}$ as an additive $(n,4^{k},d)$ code.
Decompose $H_Q$ into its $Z$-matrix and $X$-matrix as $H_Q =  H_Z + \omega H_X$
and define $2(n-k)\times(n-k)$ binary matrices $A_Z$ and $A_X$ to be their first $n-k$ columns
and $2(n-k)\times k$ binary matrices $N_Z$ and $N_X$ to be their remaining $k$ columns so that
\[H_Z = \left[\begin{array}{cc}A_Z & N_Z\end{array}\right]\]
and
\[H_X = \left[\begin{array}{cc}A_X & N_X\end{array}\right].\]
Let $A = \left[\begin{array}{cc} A_Z  & A_X \end{array}\right]$ be the $2(n-k)\times2(n-k)$ binary square matrix obtained by placing $A_X$ to the right of $A_Z$.
Note that because the square submatrix formed by the first $n-k$ columns of $H$ is full rank over $\mathbb{F}_4$,
the square matrix $A$ is full rank over $\mathbb{F}_2$.

Let $\ket{0}_X = \frac{\ket{0}+\ket{1}}{\sqrt{2}}$ and $\ket{1}_X = \frac{\ket{0}-\ket{1}}{\sqrt{2}}$,
where $\ket{0}$ and $\ket{1}$ are the computational basis.
We use $2(n-k)$ qubits in the joint $+1$ eigenstate $\ket{0}^{\otimes 2(n-k)}_X$ of $X^{\otimes 2(n-k)}$ as auxiliary qubits.
In what follows, a tensor product in this basis is labeled by a column vector
as opposed to by a row vector, which is conventional in the computational basis,
so that we write $\ket{(a_0,\dots,a_{n-1})^T}_X$ to mean $\ket{a_0}_X\otimes\dots\otimes\ket{a_{n-1}}_X$, where $a_i \in \mathbb{F}_2$ for $0 \leq i \leq n-1$.

For a unitary operator $U$ and a binary vector $\boldsymbol{a} = (a_0, \dots, a_{k-1}) \in \mathbb{F}_2^k$,
define $U^{\boldsymbol{a}}$ as the $k$-fold tensor product $O_0 \otimes \dots\otimes O_{k-1}$, where
$O_i = U$ if $a_i = 1$ and $O_i$ is the identity operator otherwise.
Take an arbitrary $k$-qubit state $\ket{\psi}$ to be encoded and $2(n-k)$ auxiliary qubits $\ket{0}^{\otimes 2(n-k)}_X$.
We use the following encoding operator
\begin{align*}
Q = \sum_{\mu \in \mathbb{F}_2^{2(n-k)}}
\ket{\mu A}
\bra{\mu A}
\otimes X^{\mu N_X}Z^{\mu N_Z}
\end{align*}
so that the encoded state is $Q\ket{0}^{\otimes 2(n-k)}_X\ket{\psi}$ that consists of $2n-k$ physical qubits.
We show that decoding operator $Q^\dag$ allows for identifying bit errors and phase errors that occur on the encoded state
as long as the number of erroneous qubits is less than or equal to $\left\lfloor\frac{d-1}{2}\right\rfloor$.
As we will see, our decoding method is a quantum variant of syndrome decoding that discretizes noise.
Hence, it is enough to correct errors due to Pauli operators $X$, $Z$, and both at the same time to be able to correct an arbitrary general error
expressed by a linear combination of $I$, $X$, $Y$, and $Z$.

We let binary vectors represent which types of error occurred on which qubits.
Take a pair $\boldsymbol{e}_X, \boldsymbol{e}_Z \in \mathbb{F}_2^{2n-k}$ of $(2n-k)$-dimensional vectors.
Define ${\boldsymbol{e}_X}_l$ and ${\boldsymbol{e}_X}_r$ as
the first $2(n-k)$ and the remaining $k$ bits of ${\boldsymbol{e}_X}$ respectively so that $\boldsymbol{e}_X = ({\boldsymbol{e}_X}_l, {\boldsymbol{e}_X}_r)$.
Define similarly $\boldsymbol{e}_Z = ({{\boldsymbol{e}_Z}_l}_0, {{\boldsymbol{e}_Z}_l}_1, {\boldsymbol{e}_Z}_r)$,
where ${{\boldsymbol{e}_Z}_l}_0$, ${{\boldsymbol{e}_Z}_l}_1$, and ${\boldsymbol{e}_Z}_r$ are
the first $n-k$, the next $n-k$, and the last $k$ bits of ${\boldsymbol{e}_Z}$ respectively.
We assume that a bit error occurred on the $i$th qubit if and only if the $i$th entry of $\boldsymbol{e}_X$ is $1$.
The location of each phase error is specified by $\boldsymbol{e}_Z$ the same way.
If the $i$th entries of $\boldsymbol{e}_X$ and $\boldsymbol{e}_Z$ are both $1$, it indicates that the $Y$ operator acted on the corresponding qubit.
Note that the first $2(n-k)$ qubits are the auxiliary ones $\ket{0}^{\otimes 2(n-k)}_X$.
The assumption that only phase errors are allowed on auxiliary qubits dictates that ${\boldsymbol{e}_X}_l$ be the zero vector $\boldsymbol{0}$.
In this case, the $n$-dimensional vector
$\boldsymbol{e} = \omega^2({\boldsymbol{e}_Z}_{l_0}, {\boldsymbol{e}_X}_r) + ({\boldsymbol{e}_Z}_{l_1}, {\boldsymbol{e}_Z}_r) \in \mathbb{F}_4^n$
contains all information about the types and locations of errors.
The correspondence between each bit of the binary components of $\boldsymbol{e}$ and the type and location of each error is summarized in Figure \ref{figcorrespondence}.

\setlength{\unitlength}{8.4mm}
\begin{figure}
\centering
\begin{picture}(10,0.7)
\put(0,0){{\small $({\boldsymbol{e}_Z}_{l_0}, {\boldsymbol{e}_X}_r) = (e_0,\dots, e_{n-k-1}\, \vert\, e_{n-k},\dots, e_{n-1})$}}
\end{picture}

\begin{picture}(9.7,1.5)
\put(0.8,0.2){\vector(2,1){2}}
\put(2.2,0.4){{\small $Z$ errors}}

\put(6,0.2){\vector(0,1){1}}
\put(6.3,0.4){{\small $X$ errors}}

\put(0.3,-0.8){$\overbrace{\phantom{llllllc}}^\text{$n-k$}\phantom{\underbrace{\phantom{lLaa}}_\text{$n-k$}}
\overbrace{\phantom{lllllllllllllllllllllllllllllllllllllllllll000000lllllllllll}}^\text{$k$}$}

\color[rgb]{0.8,0.8,0.8}
\put(2.1,-1.2){\rule{2.52mm}{4.2mm}}
\put(2.4,-1.2){\rule{2.52mm}{4.2mm}}
\put(2.7,-1.2){\rule{2.52mm}{4.2mm}}
\put(3,-1.2){\rule{2.52mm}{4.2mm}}
\put(3.3,-1.2){\rule{2.52mm}{4.2mm}}
\put(3.6,-1.2){\rule{2.52mm}{4.2mm}}
\put(3.9,-1.2){\rule{2.52mm}{4.2mm}}
\put(4.2,-1.2){\rule{2.52mm}{4.2mm}}
\put(4.5,-1.2){\rule{2.52mm}{4.2mm}}
\put(4.8,-1.2){\rule{2.52mm}{4.2mm}}
\put(5.1,-1.2){\rule{2.52mm}{4.2mm}}
\put(5.4,-1.2){\rule{2.52mm}{4.2mm}}
\put(5.7,-1.2){\rule{2.52mm}{4.2mm}}
\put(6,-1.2){\rule{2.52mm}{4.2mm}}
\put(6.3,-1.2){\rule{2.52mm}{4.2mm}}
\put(6.6,-1.2){\rule{2.52mm}{4.2mm}}
\put(6.9,-1.2){\rule{2.52mm}{4.2mm}}
\put(7.2,-1.2){\rule{2.52mm}{4.2mm}}
\put(7.5,-1.2){\rule{2.52mm}{4.2mm}}
\put(7.8,-1.2){\rule{2.52mm}{4.2mm}}
\put(8.1,-1.2){\rule{2.52mm}{4.2mm}}
\put(8.4,-1.2){\rule{2.52mm}{4.2mm}}
\put(8.7,-1.2){\rule{2.52mm}{4.2mm}}
\put(9,-1.2){\rule{2.52mm}{4.2mm}}
\put(9.3,-1.2){\rule{2.52mm}{4.2mm}}
\put(9.6,-1.2){\rule{2.52mm}{4.2mm}}
\color{black}

\put(0.3,-1.2){\line(1,0){9.6}}
\put(0.3,-0.7){\line(1,0){9.6}}

\put(0.3,-1.2){\line(0,1){0.5}}
\put(0.6,-1.2){\line(0,1){0.5}}
\put(0.9,-1.2){\line(0,1){0.5}}
\put(1.2,-1.2){\line(0,1){0.5}}
\put(1.5,-1.2){\line(0,1){0.5}}
\put(1.8,-1.2){\line(0,1){0.5}}
\put(2.1,-1.2){\line(0,1){0.5}}
\put(2.4,-1.2){\line(0,1){0.5}}
\put(2.7,-1.2){\line(0,1){0.5}}
\put(3,-1.2){\line(0,1){0.5}}
\put(3.3,-1.2){\line(0,1){0.5}}
\put(3.6,-1.2){\line(0,1){0.5}}
\put(3.9,-1.2){\line(0,1){0.5}}
\put(4.2,-1.2){\line(0,1){0.5}}
\put(4.5,-1.2){\line(0,1){0.5}}
\put(4.8,-1.2){\line(0,1){0.5}}
\put(5.1,-1.2){\line(0,1){0.5}}
\put(5.4,-1.2){\line(0,1){0.5}}
\put(5.7,-1.2){\line(0,1){0.5}}
\put(6,-1.2){\line(0,1){0.5}}
\put(6.3,-1.2){\line(0,1){0.5}}
\put(6.6,-1.2){\line(0,1){0.5}}
\put(6.9,-1.2){\line(0,1){0.5}}
\put(7.2,-1.2){\line(0,1){0.5}}
\put(7.5,-1.2){\line(0,1){0.5}}
\put(7.8,-1.2){\line(0,1){0.5}}
\put(8.1,-1.2){\line(0,1){0.5}}
\put(8.4,-1.2){\line(0,1){0.5}}
\put(8.7,-1.2){\line(0,1){0.5}}
\put(9,-1.2){\line(0,1){0.5}}
\put(9.3,-1.2){\line(0,1){0.5}}
\put(9.6,-1.2){\line(0,1){0.5}}
\put(9.9,-1.2){\line(0,1){0.5}}

\put(0.28,-1.4){$\phantom{\overbrace{\phantom{llll00}}_\text{$n-k$}}\underbrace{\phantom{lllllll}}_\text{$n-k$}
\underbrace{\phantom{lllllllllllllllllllllllllllllllllllllllllll000000lllllllllll}}_\text{$k$}$}

\put(6,-2.1){\vector(0,-1){1}}
\put(6.3,-2.7){{\small $Z$ errors}}
\put(1.65,-2.1){\vector(1,-1){0.96}}
\put(2.6,-2.7){{\small $Z$ errors}}
\end{picture}

\begin{picture}(10,3.6)
\put(0,0){{\small $({\boldsymbol{e}_Z}_{l_1}, {\boldsymbol{e}_Z}_r) = (e'_0,\dots, e'_{n-k-1}\, \vert\, e'_{n-k},\dots, e'_{n-1})$}}
\end{picture}
\caption{Correspondence of errors to binary components of $\boldsymbol{e}$. $\vert$
The white boxes represent the $2(n-k)$ less noisy qubits that may experience only phase errors.
The gray boxes are the $k$ noisy qubits that may suffer from bit errors, phase errors, or both.
The pair ${\boldsymbol{e}_Z}_{l_0}$, ${\boldsymbol{e}_Z}_{l_1}$ of the first $n-k$ bits correspond to whether phase errors occurred on the $2(n-k)$ less noisy qubits.
The $k$ bits ${\boldsymbol{e}_X}_r$ indicate whether bit errors occurred on the $k$ noisy qubits
while the other remaining $k$ bits ${\boldsymbol{e}_Z}_r$ correspond to phase errors on these noisy qubits.}\label{figcorrespondence}
\end{figure}
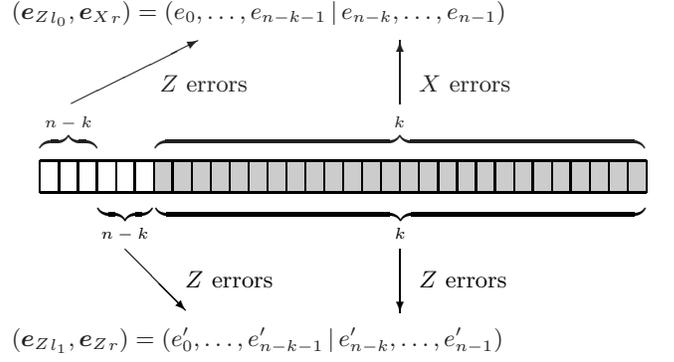

The following lemma provides the foundation of our method for extracting information about the types and locations of errors
as the trace syndrome $\operatorname{Tr}\mkern-\medmuskip\left(H_Q\boldsymbol{e}^T\right)$.

\begin{lemma}\label{qeclemma}
Let $H_Q$ be a trace parity-check matrix of a linear $[n,k,d]_4$ code.
Let $H_Z = \left[\begin{array}{cc}A_Z & N_Z\end{array}\right]$ and $H_X = \left[\begin{array}{cc}A_X & N_X\end{array}\right]$
be the $Z$-matrix and $X$-matrix of $H_Q$ respectively such that $A = \left[\begin{array}{cc} A_Z  & A_X \end{array}\right]$ is a square matrix of full rank.
Define a unitary operator
\begin{align*}\label{encoder}
Q = \sum_{\mu \in \mathbb{F}_2^{2(n-k)}}
\ket{\mu A}
\bra{\mu A}
\otimes X^{\mu N_X}Z^{\mu N_Z}.
\end{align*}
Take an arbitrary $k$-qubit state $\ket{\psi}$ and $(2n-k)$-dimensional vectors
$\boldsymbol{e}_X = ({\boldsymbol{e}_X}_l, {\boldsymbol{e}_X}_r)$
and $\boldsymbol{e}_Z = ({{\boldsymbol{e}_Z}_l}_0, {{\boldsymbol{e}_Z}_l}_1, {\boldsymbol{e}_Z}_r)$,
where ${\boldsymbol{e}_X}_l \in \mathbb{F}_2^{2(n-k)}$,
${\boldsymbol{e}_Z}_{l_0}, {\boldsymbol{e}_Z}_{l_1} \in \mathbb{F}_2^{n-k}$, and ${\boldsymbol{e}_Z}_r, {\boldsymbol{e}_X}_r \in \mathbb{F}_2^k$.
Let $\boldsymbol{e} = \omega^2({\boldsymbol{e}_Z}_{l_0}, {\boldsymbol{e}_X}_r) + ({\boldsymbol{e}_Z}_{l_1}, {\boldsymbol{e}_Z}_r)$.
Then
\begin{align*}
Q^{\dag}X^{\boldsymbol{e}_X}&Z^{\boldsymbol{e}_Z}Q\ket{0}^{\otimes 2(n-k)}_X\ket{\psi}\\
&= \ket{A^{-1}\left(\operatorname{Tr}\mkern-\medmuskip\left(H_Q\boldsymbol{e}^T\right) + N_Z{N_X}^T{\left({\boldsymbol{e}_X}_l A^{-1}\right)}^T\right)}_X\\
&\quad\otimes X^{{\boldsymbol{e}_X}_l A^{-1}N_X + {\boldsymbol{e}_X}_r}Z^{{\boldsymbol{e}_X}_l A^{-1}N_Z + {\boldsymbol{e}_Z}_r}\ket{\psi}.
\end{align*}
\end{lemma}
\Proof
Because $A$ is full rank, it has the inverse $A^{-1}$ over $\mathbb{F}_2$.
By the same token, the completeness relation asserts that
\[\sum_{\mu \in \mathbb{F}_2^{2(n-k)}}\ket{\mu A}\bra{\mu A} = I.\]
A linear algebraic computation shows that
\begin{widetext}
\begin{align}\label{longeq}
\begin{split}
Q^{\dag}&X^{\boldsymbol{e}_X}Z^{\boldsymbol{e}_Z}Q\ket{0}^{\otimes 2(n-k)}_X\ket{\psi}\\
&= \ket{A^{-1}\left(A{{\boldsymbol{e}_Z}_l}^T + N_Z{{\boldsymbol{e}_X}_r}^T + N_X{{\boldsymbol{e}_Z}_r}^T + N_Z{N_X}^T{\left({\boldsymbol{e}_X}_l A^{-1}\right)}^T\right)}_X
\otimes X^{{\boldsymbol{e}_X}_l A^{-1}N_X + {\boldsymbol{e}_X}_r}Z^{{\boldsymbol{e}_X}_l A^{-1}N_Z + {\boldsymbol{e}_Z}_r}\ket{\psi}
\end{split}
\end{align}
\end{widetext}
(see Appendix \ref{derivation} for a step-by-step derivation).
By Proposition \ref{tracesyndromeexpression}, we have
\begin{align*}
\operatorname{Tr}\mkern-\medmuskip\left(H_Q\boldsymbol{e}^T\right) &= H_Z\textup{Tr}\mkern-\medmuskip\left(\boldsymbol{e}^T\right)
+H_X\textup{Tr}\mkern-\medmuskip\left(\omega\boldsymbol{e}^T\right)\\
&= A({{\boldsymbol{e}_Z}_l}_0, {{\boldsymbol{e}_Z}_l}_1)^T + N_Z{{\boldsymbol{e}_X}_r}^T + N_X{{\boldsymbol{e}_Z}_r}^T\\
&= A{\boldsymbol{e}_Z}_l + N_Z{{\boldsymbol{e}_X}_r}^T + N_X{{\boldsymbol{e}_Z}_r}^T.
\end{align*}
Plugging the above equation into Equation (\ref{longeq}) gives the desired equation.
\qed

Recall that the assumption that auxiliary qubits are only subject to phase errors corresponds to the condition that ${\boldsymbol{e}_X}_l = \boldsymbol{0}$.
In this case, Lemma \ref{qeclemma} ensures that
\begin{align*}
Q^{\dag}X^{\boldsymbol{e}_X}Z^{\boldsymbol{e}_Z}Q&\ket{0}^{\otimes 2(n-k)}_X\ket{\psi}\\
&= \ket{A^{-1}\operatorname{Tr}\mkern-\medmuskip\left(H_Q\boldsymbol{e}^T\right)}_X
\otimes X^{{\boldsymbol{e}_X}_r}Z^{{\boldsymbol{e}_Z}_r}\ket{\psi}.
\end{align*}
Hence, measuring the auxiliary qubits and multiplying the outcome by $A$ from the left reveal the trace syndrome $\operatorname{Tr}(H_Q\boldsymbol{e}^T)$.
Because $\operatorname{wt}(\boldsymbol{e})$ is the number of erroneous qubits,
Proposition \ref{syndromedecoding} guarantees that any combination of errors on up to $\left\lfloor\frac{d-1}{2}\right\rfloor$ qubits can be corrected,
showing that the error correction capability of the linear code is fully converted to the quantum one.
Note that in Lemma \ref{qeclemma} the trace parity-check matrix can be chosen arbitrarily as long as $A$ is full rank.
Equation (\ref{tracepcm}) ensures that any parity-check matrix of any linear code over $\mathbb{F}_4$ can be directly used.
We state the result of the argument in the form of a theorem below.
\begin{theorem}\label{lessnoisyquaternary}
Let $H$ be a parity-check matrix of a linear $[n,k,d]_4$ code in which the first $n-k$ columns are linearly independent
and $H_Q$ the trace parity-check matrix whose rows are $H$ and $\omega H$.
Assume that $2n-k$ physical qubits $q_i$, $0 \leq i \leq 2n-k-1$, are sent through a noisy quantum channel in which
the first $2(n-k)$ qubits $q_i$, $0 \leq i \leq 2(n-k)-1$ are only subject to phase errors while
the remaining $k$ qubits $q_i$, $2(n-k) \leq i \leq 2n-k-1$ are subject to both bit errors and phase errors.
Define a pair $({\boldsymbol{e}_Z}_{l_0}, {\boldsymbol{e}_X}_r) = (e_0,\dots,e_{n-1}),
({\boldsymbol{e}_Z}_{l_1}, {\boldsymbol{e}_Z}_r) = (e'_0,\dots,e'_{n-1}) \in \mathbb{F}_2^{n}$ of $n$-dimensional vectors
such that for $0 \leq i \leq n-k-1$, $e_i = 1$ if a phase error occurred on $q_i$ and $e_i = 0$ otherwise, such that for $n-k \leq i \leq n-1$,
$e_i = 1$ if a bit error occurred on $q_{i+n-k}$ and $e_i = 0$ otherwise,
and such that for $0 \leq i \leq n-1$, $e'_{i} = 1$ if a phase error occurred on $q_{i+n-k}$ and $e'_i = 0$ otherwise.
Define $\boldsymbol{e} = \omega^2({\boldsymbol{e}_Z}_{l_0}, {\boldsymbol{e}_X}_r) + ({\boldsymbol{e}_Z}_{l_1}, {\boldsymbol{e}_Z}_r)$.
There exist an encoding operation that encodes logical $k$ qubits into $2n-k$ physical qubits
and a decoding operation that extracts classical information about errors
as $(n-k)$-dimensional vector $\operatorname{Tr}\mkern-\medmuskip\left(H_Q\boldsymbol{e}^T\right)$, thereby allowing
for correcting any combination of errors on up to $\left\lfloor\frac{d-1}{2}\right\rfloor$ qubits.
\end{theorem}

In the remainder of this section, we consider the case when binary linear codes are used to construct quantum error-correcting codes assisted by a phase damping channel.
Let $H_Z = \left[\begin{array}{cc}A_Z&B_Z\end{array}\right]$ and $H_X = \left[\begin{array}{cc}A_X&B_X\end{array}\right]$
be full rank parity-check matrices of linear $[n_0,k,d_0]$ code $\mathcal{C}_0$ and linear $[n_1,k,d_1]$ code $\mathcal{C}_1$ respectively,
where $A_Z$, $A_X$, $B_Z$, and $B_X$ are $(n_0-k) \times (n_0-k)$, $(n_1-k) \times (n_1-k)$, $(n_0-k) \times k$, and $(n_1-k) \times k$ matrices respectively.
Without loss of generality, we assume that $A_Z$ and $A_X$ are both full rank.
Define $(n_0+n_1 - 2k)\times(n_0+n_1 - 2k)$ binary square matrix $A$ and $(n_0+n_1 - 2k)\times k$ binary matrices $N_Z$, $N_X$ to be
\begin{align*}
A &= \left[\begin{array}{cc}A_Z & 0\\ 0 & A_X\end{array}\right],\\
N_Z &= \left[\begin{array}{c}B_Z\\ 0\end{array}\right],
\end{align*}
and
\[N_X = \left[\begin{array}{c}0\\ B_X\end{array}\right]\]
respectively.

We use the following encoding operator $Q_b$
\begin{align}\label{binaryencoder}
Q_b = \sum_{\mu \in \mathbb{F}_2^{n_0+n_1-2k}}
\ket{\mu A}
\bra{\mu A}
\otimes X^{\mu N_X}Z^{\mu N_Z}
\end{align}
analogous to the quaternary case.
Define error vectors $\boldsymbol{e}_X = ({\boldsymbol{e}_X}_l, {\boldsymbol{e}_X}_r),
\boldsymbol{e}_Z = ({{\boldsymbol{e}_Z}_l}_0, {{\boldsymbol{e}_Z}_l}_1, {\boldsymbol{e}_Z}_r) \in \mathbb{F}_2^{n_0+n_1-k}$ analogously so that
${{\boldsymbol{e}_Z}_l}_0 \in \mathbb{F}_2^{n_0-k}$,
${{\boldsymbol{e}_Z}_l}_1 \in \mathbb{F}_2^{n_1-k}$,
${\boldsymbol{e}_X}_l = \mathbb{F}_2^{n_0+n_1-k}$, and
${\boldsymbol{e}_Z}_r, {\boldsymbol{e}_X}_r \in \mathbb{F}_2^{k}$.
Take an arbitrary $k$-qubit state $\ket{\psi}$. Following the same argument as in the quaternary case,
a routine calculation proves that if the first $n_0+n_1-k$ qubits may only suffer from phase errors, we have
\begin{align*}
Q^{\dag}_bX^{\boldsymbol{e}_X}Z^{\boldsymbol{e}_Z}&Q_b\ket{0}^{\otimes n_0+n_1-2k}_X\ket{\psi}\\
&= \ket{A_Z^{-1}H_Z({{\boldsymbol{e}_Z}_l}_0, {\boldsymbol{e}_X}_r)^T}_X\\
&\quad\otimes\ket{A_X^{-1}H_X({{\boldsymbol{e}_Z}_l}_1, {\boldsymbol{e}_Z}_r)^T}_X
X^{{\boldsymbol{e}_X}_r}Z^{{\boldsymbol{e}_Z}_r}\ket{\psi}.
\end{align*}
Measuring the first $n_0+n_1-k$ auxiliary qubits and multiplying the outcome by $A_Z$ from the left gives
the binary vector $H_Z({{\boldsymbol{e}_Z}_l}_0, {\boldsymbol{e}_X}_r)^T$.
Because $H_Z$ is a parity-check matrix of the linear $[n_0,k,d_0]$ code $\mathcal{C}_0$,
the encoding operator $Q_b$ given in Equation (\ref{binaryencoder}) and the corresponding decoding operator $Q^{\dag}_b$ can identify
the vector $({{\boldsymbol{e}_Z}_l}_0, {\boldsymbol{e}_X}_r)$ from the syndrome $H_Z({{\boldsymbol{e}_Z}_l}_0, {\boldsymbol{e}_X}_r)^T$
under the assumption that $\operatorname{wt}(({{\boldsymbol{e}_Z}_l}_0, {\boldsymbol{e}_X}_r)) \leq \left\lfloor\frac{d_0-1}{2}\right\rfloor$.
By the same token, the syndrome $H_X({{\boldsymbol{e}_Z}_l}_1, {\boldsymbol{e}_Z}_r)^T$ obtained from the next $n_0+n_1-k$ auxiliary qubits
reveals the vector $({{\boldsymbol{e}_Z}_l}_1, {\boldsymbol{e}_Z}_r)$ under the assumption that
$\operatorname{wt}(({{\boldsymbol{e}_Z}_l}_1, {\boldsymbol{e}_Z}_r)) \leq \left\lfloor\frac{d_1-1}{2}\right\rfloor$.

If $\mathcal{C}_0 = \mathcal{C}_1$, the pair $Q_b$, $Q^{\dag}_b$ of operations can correct any combination of errors on up to
$\left\lfloor\frac{d_0-1}{2}\right\rfloor = \left\lfloor\frac{d_1-1}{2}\right\rfloor$ qubits.
Because a parity-check matrix of a linear code of length $n$ and dimension $k$ always contains a set of $n-k$ linearly independent columns,
any parity-check matrix of any binary linear code can be used this way.

One may also tailor the error correction capabilities to suppress asymmetric noise more efficiently
using fewer physical qubits by employing two linear codes with different minimum distances.
If we fix the dimension $k$,
the smallest $n$ such that there exits a linear $[n,k,d]$ code generally becomes smaller as $d$ becomes smaller. 
Hence, if one linear code is not required to be as strong as the other due to asymmetry in noise,
the number $n_0+n_1-k$ of physical qubits required to protect $k$ logical qubits can be smaller.

We conclude this section by summarizing the result in the binary case as a theorem.
\begin{theorem}\label{lessnoisybinary}
Let $H_Z$ and $H_X$ be parity-check matrices of linear codes of parameters $[n_0,k,d_0]$ and $[n_1,k,d_1]$ respectively,
where the first $n_0-k$ and first $n_1-k$ columns are both linearly independent.
Assume that $n_0+n_1-k$ physical qubits $q_i$, $0 \leq i \leq n_0+n_1-k-1$, are sent through a noisy quantum channel in which
the first $n_0+n_1-2k$ qubits $q_i$, $0 \leq i \leq n_0+n_1-2k-1$ are only subject to phase errors while
the remaining $k$ qubits $q_i$, $n_0+n_1-2k \leq i \leq n_0+n_1-k-1$ are subject to both bit errors and phase errors.
Define $({\boldsymbol{e}_Z}_{l_0}, {\boldsymbol{e}_X}_r) = (e_0,\dots,e_{n-1}) \in \mathbb{F}_2^{n_0}$ and
$({\boldsymbol{e}_Z}_{l_1}, {\boldsymbol{e}_Z}_r) = (e'_0,\dots,e'_{n-1}) \in \mathbb{F}_2^{n_1}$,
where ${{\boldsymbol{e}_Z}_l}_0 \in \mathbb{F}_2^{n_0-k}$,
${{\boldsymbol{e}_Z}_l}_1 \in \mathbb{F}_2^{n_1-k}$, and
${\boldsymbol{e}_Z}_r, {\boldsymbol{e}_X}_r \in \mathbb{F}_2^{k}$
such that for $0 \leq i \leq n_0-k-1$, $e_i = 1$ if a phase error occurred on $q_i$ and $e_i = 0$ otherwise, such that for $n_0-k \leq i \leq n_0-1$,
$e_i = 1$ if a bit error occurred on $q_{i+n_1-k}$ and $e_i = 0$ otherwise,
and such that for $0 \leq i \leq n_1-1$, $e'_{i} = 1$ if a phase error occurred on $q_{i+n_0-k}$ and $e'_i = 0$ otherwise.
There exist an encoding operation that encodes logical $k$ qubits into $n_0+n_1-k$ physical qubits
and a decoding operation that extracts classical information about errors
as $(n_0-k)$-dimensional vector $H_Z({\boldsymbol{e}_Z}_{l_0}, {\boldsymbol{e}_X}_r)^T$
and $(n_1-k)$-dimensional vector $H_X({\boldsymbol{e}_Z}_{l_1}, {\boldsymbol{e}_Z}_r)^T$,
thereby allowing
for correcting any combination of errors under the assumption that
$\operatorname{wt}(({\boldsymbol{e}_Z}_{l_0}, {\boldsymbol{e}_X}_r)) \leq \left\lfloor\frac{d_1-1}{2}\right\rfloor$
and $\operatorname{wt}(({\boldsymbol{e}_Z}_{l_1}, {\boldsymbol{e}_Z}_r)) \leq \left\lfloor\frac{d_1-1}{2}\right\rfloor$.
\end{theorem}

\section{Concluding remarks}
We have generalized the encoding and decoding operations for quantum error correction assisted by less noisy qubits
in such a way that any parity-check matrix of any binary or quaternary linear code can be imported into the quantum domain.
This is achieved while the active error correction capability of the underlying classical code is still fully and directly utilized for quantum error correction
through the principle of error syndromes in classical coding theory.
Hence, our generalization eliminated the weakness of the original formulation of the assisted scheme
that parity-check matrices were required to be of particular form.

To see the benefit of being able to use any parity-check matrix of any linear code, consider the sum-product algorithm for error correction,
which can efficiently infer errors from syndromes (see \cite{MacKay:2003} for details of the algorithm).
As mentioned earlier, this algorithm is usable only if parity-check matrices are of special sparse form.
Linear codes that admit parity-check matrices suitable for this decoding method are called \textit{low-density parity-check} (LDPC) \textit{codes}.
Quantum error-correcting codes that use the sum-product algorithm in the same manner are called \textit{quantum LDPC codes}
(see, for example, \cite{MacKay:2004,Poulin:2008,Hsieh:2009}).
Because our generalized assisted scheme directly uses parity-check matrices and syndromes in the same way as in classical coding theory,
any binary LDPC code can be turned into a quantum LDPC code without losing its suitability for the sum-product algorithm.

To give a concrete example, consider the linear $[1080,999,6]$ code from affine geometry AG$(4,3)$, which produces
an entanglement-assisted quantum LDPC code that encodes $998$ logical qubits into $1080$ physical qubits
assisted by $80$ maximally entangled noiseless qubits \cite{Fujiwara:2010e}.
With assistance from less noisy qubits, the corresponding quantum error-correcting code encodes $999$ logical qubits into $1161$ physical qubits
of which $162$ are assumed to be free from bit errors.
The underlying linear code belongs to the class of LDPC codes that have been studied
for very high-rate information transmission in the classical domain \cite{Vasic:2002,Johnson:2003,Vasic:2004,Ammar:2004}.
Because the entanglement-assisted stabilizer formalism can directly exploit the parity-check matrix in sparse form,
the entanglement-assisted quantum LDPC code performs as well as the classical one used as its ingredient.
If we employ the original formulation of assistance by less noisy qubits to relax the burden of providing noiseless auxiliary qubits, however,
the parity-check matrix must be modified to obtain a similar one in standard form.
Simulations performed in \cite{Fujiwara:2014} observed a noticeable drop in error correction performance due to this modification under the sum-product algorithm.
The results presented in this paper eliminated the need of modification,
which means that assistance from less noisy qubits, rather than by maximally entangled noiseless ones,
is enough to reproduce in the quantum domain
the performance of the linear $[1080,999,6]$ code as a high-rate LDPC code.

Another improvement we made is that our framework can now flexibly construct quantum error-correcting codes
that have different degrees of tolerance against different types of noise.
We made this possible by showing that two different binary linear codes can be used to create one quantum error-correcting code.
It is notable that a reduction in the number of physical qubits of the encoded state is achieved
by importing a pair of linear codes that have different lengths but are of the same dimension.
This is in contrast to the typical construction technique for asymmetric quantum error-correcting codes through the Calderbank-Shor-Steane (CSS) construction,
where two linear codes of the same length that have different dimensions are used \cite{Calderbank:1996,Steane:1996}.

It should be noted, however, that each binary linear code is responsible for correcting errors on both noisy and less noisy qubits.
Hence, optimization to asymmetric noise is tricker than if there are only noisy qubits.
While we do not expect that this would cause a major issue because the dominant source of errors would be the fully noisy channel,
it is a weakness of our framework that designing a perfectly optimized code will likely require a highly technical construction.

One aspect we did not explicitly mention is possible use of parity-check matrices with redundant rows.
A parity-check matrix of a linear code is usually assumed to be full rank and has no linearly dependent rows in classical coding theory.
In fact, from the viewpoint of syndrome decoding, such extra rows do not provide any further information about errors.
Nonetheless, it is known that having redundant rows can be beneficial when a suboptimal decoding method is used \cite{MacKay:2003}.
Our generalized quantum error-correcting codes can take advantage of this phenomenon as well.
This is because any extra bit from a redundant row in the extended syndrome can be expressed as a linear combination of bits of the syndrome
from a parity-check matrix of full rank.
Hence, if redundant syndrome bits are required to aid a suboptimal decoder,
one may compute them after measuring auxiliary qubits.

Regarding a possible major disadvantage of our encoding and decoding procedures,
it should be noted that the number of auxiliary qubits required for our coding scheme is solely determined by the parameters of the underlying linear code.
This is fortunate if one would like to import a linear code of large information rate because then
it is guaranteed that only a small fraction of physical qubits need to be less noisy.
However, this is a double-edged sword because if one wishes to employ a linear code of low rate,
the number of required auxiliary qubits is always large, which can render our idea unusuitable in such situations.
We hope to develop encoding and decoding procedures that improve on our current approach in this regard in future work.

Another important question we did not address in this work is the theoretical limit of noise suppression achievable by assistance from less noisy qubits.
We focused on how to import classical error-correcting codes into the quantum domain.
This point of view is useful when explicit constructions for quantum error-correcting codes are of concern.
However, it is also quite important to investigate the concept of assisted quantum error correction more generally from a Shannon theoretic viewpoint.
We hope that research from different viewpoints will give greater insight into assisted quantum error correction.


\begin{thebibliography}{30}%
\makeatletter
\providecommand \@ifxundefined [1]{%
 \@ifx{#1\undefined}
}%
\providecommand \@ifnum [1]{%
 \ifnum #1\expandafter \@firstoftwo
 \else \expandafter \@secondoftwo
 \fi
}%
\providecommand \@ifx [1]{%
 \ifx #1\expandafter \@firstoftwo
 \else \expandafter \@secondoftwo
 \fi
}%
\providecommand \natexlab [1]{#1}%
\providecommand \enquote  [1]{``#1''}%
\providecommand \bibnamefont  [1]{#1}%
\providecommand \bibfnamefont [1]{#1}%
\providecommand \citenamefont [1]{#1}%
\providecommand \href@noop [0]{\@secondoftwo}%
\providecommand \href [0]{\begingroup \@sanitize@url \@href}%
\providecommand \@href[1]{\@@startlink{#1}\@@href}%
\providecommand \@@href[1]{\endgroup#1\@@endlink}%
\providecommand \@sanitize@url [0]{\catcode `\\12\catcode `\$12\catcode
  `\&12\catcode `\#12\catcode `\^12\catcode `\_12\catcode `\%12\relax}%
\providecommand \@@startlink[1]{}%
\providecommand \@@endlink[0]{}%
\providecommand \url  [0]{\begingroup\@sanitize@url \@url }%
\providecommand \@url [1]{\endgroup\@href {#1}{\urlprefix }}%
\providecommand \urlprefix  [0]{URL }%
\providecommand \Eprint [0]{\href }%
\providecommand \doibase [0]{http://dx.doi.org/}%
\providecommand \selectlanguage [0]{\@gobble}%
\providecommand \bibinfo  [0]{\@secondoftwo}%
\providecommand \bibfield  [0]{\@secondoftwo}%
\providecommand \translation [1]{[#1]}%
\providecommand \BibitemOpen [0]{}%
\providecommand \bibitemStop [0]{}%
\providecommand \bibitemNoStop [0]{.\EOS\space}%
\providecommand \EOS [0]{\spacefactor3000\relax}%
\providecommand \BibitemShut  [1]{\csname bibitem#1\endcsname}%
\let\auto@bib@innerbib\@empty
\bibitem [{\citenamefont {Lidar}\ and\ \citenamefont
  {Brun}(2013)}]{Lidar:2013}%
  \BibitemOpen
  \bibinfo {editor} {\bibfnamefont {D.~A.}\ \bibnamefont {Lidar}}\ and\
  \bibinfo {editor} {\bibfnamefont {T.~A.}\ \bibnamefont {Brun}},\ eds.,\
  \href@noop {} {\emph {\bibinfo {title} {Quantum Error Correction}}}\
  (\bibinfo  {publisher} {Cambridge Univ. Press},\ \bibinfo {address} {New
  York},\ \bibinfo {year} {2013})\BibitemShut {NoStop}%
\bibitem [{\citenamefont {Gottesman}(1997)}]{Gottesman:1997}%
  \BibitemOpen
  \bibfield  {author} {\bibinfo {author} {\bibfnamefont {D.}~\bibnamefont
  {Gottesman}},\ }\emph {\bibinfo {title} {Stabilizer codes and quantum error
  correction}},\ \href@noop {} {Ph.D. thesis},\ \bibinfo  {school} {California
  Institute of Technology} (\bibinfo {year} {1997})\BibitemShut {NoStop}%
\bibitem [{\citenamefont {Brun}\ \emph {et~al.}(2006)\citenamefont {Brun},
  \citenamefont {Devetak},\ and\ \citenamefont {Hsieh}}]{Brun:2006}%
  \BibitemOpen
  \bibfield  {author} {\bibinfo {author} {\bibfnamefont {T.~A.}\ \bibnamefont
  {Brun}}, \bibinfo {author} {\bibfnamefont {I.}~\bibnamefont {Devetak}}, \
  and\ \bibinfo {author} {\bibfnamefont {M.-H.}\ \bibnamefont {Hsieh}},\
  }\href@noop {} {\bibfield  {journal} {\bibinfo  {journal} {Science}\ }\textbf
  {\bibinfo {volume} {314}},\ \bibinfo {pages} {436} (\bibinfo {year}
  {2006})}\BibitemShut {NoStop}%
\bibitem [{\citenamefont {Brun}\ \emph {et~al.}(2014)\citenamefont {Brun},
  \citenamefont {Devetak},\ and\ \citenamefont {Hsieh}}]{Brun:2014}%
  \BibitemOpen
  \bibfield  {author} {\bibinfo {author} {\bibfnamefont {T.~A.}\ \bibnamefont
  {Brun}}, \bibinfo {author} {\bibfnamefont {I.}~\bibnamefont {Devetak}}, \
  and\ \bibinfo {author} {\bibfnamefont {M.-H.}\ \bibnamefont {Hsieh}},\
  }\href@noop {} {\bibfield  {journal} {\bibinfo  {journal} {{IEEE} Trans. Inf.
  Theory}\ }\textbf {\bibinfo {volume} {60}},\ \bibinfo {pages} {3073}
  (\bibinfo {year} {2014})}\BibitemShut {NoStop}%
\bibitem [{\citenamefont {Kribs}\ \emph {et~al.}(2005)\citenamefont {Kribs},
  \citenamefont {Laflamme},\ and\ \citenamefont {Poulin}}]{Kribs:2005}%
  \BibitemOpen
  \bibfield  {author} {\bibinfo {author} {\bibfnamefont {D.}~\bibnamefont
  {Kribs}}, \bibinfo {author} {\bibfnamefont {R.}~\bibnamefont {Laflamme}}, \
  and\ \bibinfo {author} {\bibfnamefont {D.}~\bibnamefont {Poulin}},\
  }\href@noop {} {\bibfield  {journal} {\bibinfo  {journal} {Phys. Rev. Lett.}\
  }\textbf {\bibinfo {volume} {94}},\ \bibinfo {pages} {180501} (\bibinfo
  {year} {2005})}\BibitemShut {NoStop}%
\bibitem [{\citenamefont {Kribs}\ \emph {et~al.}(2006)\citenamefont {Kribs},
  \citenamefont {Laflamme}, \citenamefont {Poulin},\ and\ \citenamefont
  {Lesosky}}]{Kribs:2006}%
  \BibitemOpen
  \bibfield  {author} {\bibinfo {author} {\bibfnamefont {D.}~\bibnamefont
  {Kribs}}, \bibinfo {author} {\bibfnamefont {R.}~\bibnamefont {Laflamme}},
  \bibinfo {author} {\bibfnamefont {D.}~\bibnamefont {Poulin}}, \ and\ \bibinfo
  {author} {\bibfnamefont {M.}~\bibnamefont {Lesosky}},\ }\href@noop {}
  {\bibfield  {journal} {\bibinfo  {journal} {Quantum Inf. Comput.}\ }\textbf
  {\bibinfo {volume} {6}},\ \bibinfo {pages} {382} (\bibinfo {year}
  {2006})}\BibitemShut {NoStop}%
\bibitem [{\citenamefont {Cross}\ \emph {et~al.}(2009)\citenamefont {Cross},
  \citenamefont {Smith}, \citenamefont {Smolin},\ and\ \citenamefont
  {Zeng}}]{Cross:2009}%
  \BibitemOpen
  \bibfield  {author} {\bibinfo {author} {\bibfnamefont {A.}~\bibnamefont
  {Cross}}, \bibinfo {author} {\bibfnamefont {G.}~\bibnamefont {Smith}},
  \bibinfo {author} {\bibfnamefont {J.~A.}\ \bibnamefont {Smolin}}, \ and\
  \bibinfo {author} {\bibfnamefont {B.}~\bibnamefont {Zeng}},\ }\href@noop {}
  {\bibfield  {journal} {\bibinfo  {journal} {{IEEE} Trans. Inf. Theory}\
  }\textbf {\bibinfo {volume} {55}},\ \bibinfo {pages} {433} (\bibinfo {year}
  {2009})}\BibitemShut {NoStop}%
\bibitem [{\citenamefont {Hsieh}\ \emph {et~al.}(2007)\citenamefont {Hsieh},
  \citenamefont {Devetak},\ and\ \citenamefont {Brun}}]{Hsieh:2007}%
  \BibitemOpen
  \bibfield  {author} {\bibinfo {author} {\bibfnamefont {M.-H.}\ \bibnamefont
  {Hsieh}}, \bibinfo {author} {\bibfnamefont {I.}~\bibnamefont {Devetak}}, \
  and\ \bibinfo {author} {\bibfnamefont {T.~A.}\ \bibnamefont {Brun}},\
  }\href@noop {} {\bibfield  {journal} {\bibinfo  {journal} {Phys. Rev. A}\
  }\textbf {\bibinfo {volume} {76}},\ \bibinfo {pages} {062313} (\bibinfo
  {year} {2007})}\BibitemShut {NoStop}%
\bibitem [{\citenamefont {Shin}\ \emph {et~al.}(2011)\citenamefont {Shin},
  \citenamefont {Heo},\ and\ \citenamefont {Brun}}]{Shin:2011}%
  \BibitemOpen
  \bibfield  {author} {\bibinfo {author} {\bibfnamefont {J.}~\bibnamefont
  {Shin}}, \bibinfo {author} {\bibfnamefont {J.}~\bibnamefont {Heo}}, \ and\
  \bibinfo {author} {\bibfnamefont {T.~A.}\ \bibnamefont {Brun}},\ }\href@noop
  {} {\bibfield  {journal} {\bibinfo  {journal} {Phys. Rev. A}\ }\textbf
  {\bibinfo {volume} {84}},\ \bibinfo {pages} {062321} (\bibinfo {year}
  {2011})}\BibitemShut {NoStop}%
\bibitem [{\citenamefont {Shin}\ \emph {et~al.}(2013)\citenamefont {Shin},
  \citenamefont {Heo},\ and\ \citenamefont {Brun}}]{Shin:2013}%
  \BibitemOpen
  \bibfield  {author} {\bibinfo {author} {\bibfnamefont {J.}~\bibnamefont
  {Shin}}, \bibinfo {author} {\bibfnamefont {J.}~\bibnamefont {Heo}}, \ and\
  \bibinfo {author} {\bibfnamefont {T.~A.}\ \bibnamefont {Brun}},\ }\href@noop
  {} {\emph {\bibinfo {title} {General quantum error-correcting code with
  entanglement based on codeword stabilized quantum code}}},\ \bibinfo {type}
  {e-print}\ \bibinfo {number} {arXiv:1311.1533}\ (\bibinfo {year}
  {2013})\BibitemShut {NoStop}%
\bibitem [{\citenamefont {Hsieh}\ \emph {et~al.}(2011)\citenamefont {Hsieh},
  \citenamefont {Yen},\ and\ \citenamefont {Hsu}}]{Hsieh:2011}%
  \BibitemOpen
  \bibfield  {author} {\bibinfo {author} {\bibfnamefont {M.-H.}\ \bibnamefont
  {Hsieh}}, \bibinfo {author} {\bibfnamefont {W.-T.}\ \bibnamefont {Yen}}, \
  and\ \bibinfo {author} {\bibfnamefont {L.-Y.}\ \bibnamefont {Hsu}},\
  }\href@noop {} {\bibfield  {journal} {\bibinfo  {journal} {{IEEE} Trans. Inf.
  Theory}\ }\textbf {\bibinfo {volume} {57}},\ \bibinfo {pages} {1761}
  (\bibinfo {year} {2011})}\BibitemShut {NoStop}%
\bibitem [{\citenamefont {Fujiwara}\ \emph {et~al.}(2010)\citenamefont
  {Fujiwara}, \citenamefont {Clark}, \citenamefont {Vandendriessche},
  \citenamefont {{De Boeck}},\ and\ \citenamefont {Tonchev}}]{Fujiwara:2010e}%
  \BibitemOpen
  \bibfield  {author} {\bibinfo {author} {\bibfnamefont {Y.}~\bibnamefont
  {Fujiwara}}, \bibinfo {author} {\bibfnamefont {D.}~\bibnamefont {Clark}},
  \bibinfo {author} {\bibfnamefont {P.}~\bibnamefont {Vandendriessche}},
  \bibinfo {author} {\bibfnamefont {M.}~\bibnamefont {{De Boeck}}}, \ and\
  \bibinfo {author} {\bibfnamefont {V.~D.}\ \bibnamefont {Tonchev}},\
  }\href@noop {} {\bibfield  {journal} {\bibinfo  {journal} {Phys. Rev. A}\
  }\textbf {\bibinfo {volume} {82}},\ \bibinfo {pages} {042338} (\bibinfo
  {year} {2010})}\BibitemShut {NoStop}%
\bibitem [{\citenamefont {Lai}\ and\ \citenamefont {Brun}(2012)}]{Lai:2012}%
  \BibitemOpen
  \bibfield  {author} {\bibinfo {author} {\bibfnamefont {C.-Y.}\ \bibnamefont
  {Lai}}\ and\ \bibinfo {author} {\bibfnamefont {T.~A.}\ \bibnamefont {Brun}},\
  }\href@noop {} {\bibfield  {journal} {\bibinfo  {journal} {Phys. Rev. A}\
  }\textbf {\bibinfo {volume} {86}},\ \bibinfo {pages} {032319} (\bibinfo
  {year} {2012})}\BibitemShut {NoStop}%
\bibitem [{\citenamefont {Fujiwara}(2013)}]{Fujiwara:2013c}%
  \BibitemOpen
  \bibfield  {author} {\bibinfo {author} {\bibfnamefont {Y.}~\bibnamefont
  {Fujiwara}},\ }\href@noop {} {\bibfield  {journal} {\bibinfo  {journal}
  {Phys. Rev. Lett.}\ }\textbf {\bibinfo {volume} {110}},\ \bibinfo {pages}
  {170501} (\bibinfo {year} {2013})}\BibitemShut {NoStop}%
\bibitem [{\citenamefont {Ioffe}\ and\ \citenamefont
  {M\'{e}zard}(2007)}]{Ioffe:2007}%
  \BibitemOpen
  \bibfield  {author} {\bibinfo {author} {\bibfnamefont {L.}~\bibnamefont
  {Ioffe}}\ and\ \bibinfo {author} {\bibfnamefont {M.}~\bibnamefont
  {M\'{e}zard}},\ }\href@noop {} {\bibfield  {journal} {\bibinfo  {journal}
  {Phys. Rev. A}\ }\textbf {\bibinfo {volume} {75}},\ \bibinfo {pages} {032345}
  (\bibinfo {year} {2007})}\BibitemShut {NoStop}%
\bibitem [{\citenamefont {Sarvepalli}\ \emph {et~al.}(2009)\citenamefont
  {Sarvepalli}, \citenamefont {Klappenecker},\ and\ \citenamefont
  {Rotteler}}]{Sarvepalli:2009}%
  \BibitemOpen
  \bibfield  {author} {\bibinfo {author} {\bibfnamefont {P.~K.}\ \bibnamefont
  {Sarvepalli}}, \bibinfo {author} {\bibfnamefont {A.}~\bibnamefont
  {Klappenecker}}, \ and\ \bibinfo {author} {\bibfnamefont {M.}~\bibnamefont
  {Rotteler}},\ }\href@noop {} {\bibfield  {journal} {\bibinfo  {journal}
  {Proc. R. Soc. A}\ }\textbf {\bibinfo {volume} {465}},\ \bibinfo {pages}
  {1645} (\bibinfo {year} {2009})}\BibitemShut {NoStop}%
\bibitem [{\citenamefont {MacKay}(2003)}]{MacKay:2003}%
  \BibitemOpen
  \bibfield  {author} {\bibinfo {author} {\bibfnamefont {D.~J.~C.}\
  \bibnamefont {MacKay}},\ }\href@noop {} {\emph {\bibinfo {title} {Information
  {T}heory, {I}nference, and {L}earning {A}lgorithms}}}\ (\bibinfo  {publisher}
  {Cambridge University Press},\ \bibinfo {address} {Cambridge},\ \bibinfo
  {year} {2003})\BibitemShut {NoStop}%
\bibitem [{\citenamefont {{MacWilliams}}\ and\ \citenamefont
  {Sloane}(1977)}]{MacWilliams:1977}%
  \BibitemOpen
  \bibfield  {author} {\bibinfo {author} {\bibfnamefont {F.~J.}\ \bibnamefont
  {{MacWilliams}}}\ and\ \bibinfo {author} {\bibfnamefont {N.~J.~A.}\
  \bibnamefont {Sloane}},\ }\href@noop {} {\emph {\bibinfo {title} {The Theory
  of Error-Correcting Codes}}}\ (\bibinfo  {publisher} {North-Holland
  Publishing Company},\ \bibinfo {address} {Amsterdam},\ \bibinfo {year}
  {1977})\BibitemShut {NoStop}%
\bibitem [{\citenamefont {Huffman}\ and\ \citenamefont
  {Pless}(2003)}]{Huffman:2003b}%
  \BibitemOpen
  \bibfield  {author} {\bibinfo {author} {\bibfnamefont {W.~C.}\ \bibnamefont
  {Huffman}}\ and\ \bibinfo {author} {\bibfnamefont {V.}~\bibnamefont
  {Pless}},\ }\href@noop {} {\emph {\bibinfo {title} {Fundamentals of
  Error-Correcting Codes}}}\ (\bibinfo  {publisher} {Cambridge Univ. Press},\
  \bibinfo {address} {Cambridge},\ \bibinfo {year} {2003})\BibitemShut
  {NoStop}%
\bibitem [{\citenamefont {Nielsen}\ and\ \citenamefont
  {Chuang}(2000)}]{Nielsen:2000}%
  \BibitemOpen
  \bibfield  {author} {\bibinfo {author} {\bibfnamefont {M.~A.}\ \bibnamefont
  {Nielsen}}\ and\ \bibinfo {author} {\bibfnamefont {I.~L.}\ \bibnamefont
  {Chuang}},\ }\href@noop {} {\emph {\bibinfo {title} {Quantum {C}omputation
  and {Q}uantum {I}nformation}}}\ (\bibinfo  {publisher} {Cambridge Univ.
  Press},\ \bibinfo {address} {New York},\ \bibinfo {year} {2000})\BibitemShut
  {NoStop}%
\bibitem [{\citenamefont {Fujiwara}\ \emph {et~al.}(2014)\citenamefont
  {Fujiwara}, \citenamefont {Gruner},\ and\ \citenamefont
  {Vandendriessche}}]{Fujiwara:2014}%
  \BibitemOpen
  \bibfield  {author} {\bibinfo {author} {\bibfnamefont {Y.}~\bibnamefont
  {Fujiwara}}, \bibinfo {author} {\bibfnamefont {A.}~\bibnamefont {Gruner}}, \
  and\ \bibinfo {author} {\bibfnamefont {P.}~\bibnamefont {Vandendriessche}},\
  }\href@noop {} {\emph {\bibinfo {title} {High-rate quantum low-density
  parity-check codes assisted by reliable qubits}}},\ \bibinfo {type}
  {e-print}\ \bibinfo {number} {arXiv:1309.5587}\ (\bibinfo {year}
  {2014})\BibitemShut {NoStop}%
\bibitem [{\citenamefont {MacKay}\ \emph {et~al.}(2004)\citenamefont {MacKay},
  \citenamefont {Mitchison},\ and\ \citenamefont {McFadden}}]{MacKay:2004}%
  \BibitemOpen
  \bibfield  {author} {\bibinfo {author} {\bibfnamefont {D.~J.~C.}\
  \bibnamefont {MacKay}}, \bibinfo {author} {\bibfnamefont {G.}~\bibnamefont
  {Mitchison}}, \ and\ \bibinfo {author} {\bibfnamefont {P.~L.}\ \bibnamefont
  {McFadden}},\ }\href@noop {} {\bibfield  {journal} {\bibinfo  {journal}
  {{IEEE} Trans. Inf. Theory}\ }\textbf {\bibinfo {volume} {50}},\ \bibinfo
  {pages} {2315} (\bibinfo {year} {2004})}\BibitemShut {NoStop}%
\bibitem [{\citenamefont {Poulin}\ and\ \citenamefont
  {Chung}(2008)}]{Poulin:2008}%
  \BibitemOpen
  \bibfield  {author} {\bibinfo {author} {\bibfnamefont {D.}~\bibnamefont
  {Poulin}}\ and\ \bibinfo {author} {\bibfnamefont {Y.-J.}\ \bibnamefont
  {Chung}},\ }\href@noop {} {\bibfield  {journal} {\bibinfo  {journal} {Quantum
  Inf. Comput.}\ }\textbf {\bibinfo {volume} {8}},\ \bibinfo {pages} {987}
  (\bibinfo {year} {2008})}\BibitemShut {NoStop}%
\bibitem [{\citenamefont {Hsieh}\ \emph {et~al.}(2009)\citenamefont {Hsieh},
  \citenamefont {Brun},\ and\ \citenamefont {Devetak}}]{Hsieh:2009}%
  \BibitemOpen
  \bibfield  {author} {\bibinfo {author} {\bibfnamefont {M.-H.}\ \bibnamefont
  {Hsieh}}, \bibinfo {author} {\bibfnamefont {T.~A.}\ \bibnamefont {Brun}}, \
  and\ \bibinfo {author} {\bibfnamefont {I.}~\bibnamefont {Devetak}},\
  }\href@noop {} {\bibfield  {journal} {\bibinfo  {journal} {Phys. Rev. A}\
  }\textbf {\bibinfo {volume} {79}},\ \bibinfo {pages} {032340} (\bibinfo
  {year} {2009})}\BibitemShut {NoStop}%
\bibitem [{\citenamefont {Vasi\'{c}}\ \emph {et~al.}(2002)\citenamefont
  {Vasi\'{c}}, \citenamefont {Kurtas},\ and\ \citenamefont
  {Kuznetsov}}]{Vasic:2002}%
  \BibitemOpen
  \bibfield  {author} {\bibinfo {author} {\bibfnamefont {B.}~\bibnamefont
  {Vasi\'{c}}}, \bibinfo {author} {\bibfnamefont {E.~M.}\ \bibnamefont
  {Kurtas}}, \ and\ \bibinfo {author} {\bibfnamefont {A.~V.}\ \bibnamefont
  {Kuznetsov}},\ }\href@noop {} {\bibfield  {journal} {\bibinfo  {journal}
  {{IEEE} Trans. Magn.}\ }\textbf {\bibinfo {volume} {38}},\ \bibinfo {pages}
  {1705} (\bibinfo {year} {2002})}\BibitemShut {NoStop}%
\bibitem [{\citenamefont {Johnson}\ and\ \citenamefont
  {Weller}(2003)}]{Johnson:2003}%
  \BibitemOpen
  \bibfield  {author} {\bibinfo {author} {\bibfnamefont {S.~J.}\ \bibnamefont
  {Johnson}}\ and\ \bibinfo {author} {\bibfnamefont {S.~R.}\ \bibnamefont
  {Weller}},\ }\href@noop {} {\bibfield  {journal} {\bibinfo  {journal} {{IEEE}
  Trans. Commun.}\ }\textbf {\bibinfo {volume} {51}},\ \bibinfo {pages} {1413}
  (\bibinfo {year} {2003})}\BibitemShut {NoStop}%
\bibitem [{\citenamefont {Vasi\'{c}}\ and\ \citenamefont
  {Milenkovic}(2004)}]{Vasic:2004}%
  \BibitemOpen
  \bibfield  {author} {\bibinfo {author} {\bibfnamefont {B.}~\bibnamefont
  {Vasi\'{c}}}\ and\ \bibinfo {author} {\bibfnamefont {O.}~\bibnamefont
  {Milenkovic}},\ }\href@noop {} {\bibfield  {journal} {\bibinfo  {journal}
  {{IEEE} Trans. Inf. Theory}\ }\textbf {\bibinfo {volume} {50}},\ \bibinfo
  {pages} {1156} (\bibinfo {year} {2004})}\BibitemShut {NoStop}%
\bibitem [{\citenamefont {Ammar}\ \emph {et~al.}(2004)\citenamefont {Ammar},
  \citenamefont {Honary}, \citenamefont {Kou}, \citenamefont {Xu},\ and\
  \citenamefont {Lin}}]{Ammar:2004}%
  \BibitemOpen
  \bibfield  {author} {\bibinfo {author} {\bibfnamefont {B.}~\bibnamefont
  {Ammar}}, \bibinfo {author} {\bibfnamefont {B.}~\bibnamefont {Honary}},
  \bibinfo {author} {\bibfnamefont {Y.}~\bibnamefont {Kou}}, \bibinfo {author}
  {\bibfnamefont {J.}~\bibnamefont {Xu}}, \ and\ \bibinfo {author}
  {\bibfnamefont {S.}~\bibnamefont {Lin}},\ }\href@noop {} {\bibfield
  {journal} {\bibinfo  {journal} {{IEEE} Trans. Inf. Theory}\ }\textbf
  {\bibinfo {volume} {50}},\ \bibinfo {pages} {1257} (\bibinfo {year}
  {2004})}\BibitemShut {NoStop}%
\bibitem [{\citenamefont {Calderbank}\ and\ \citenamefont
  {Shor}(1996)}]{Calderbank:1996}%
  \BibitemOpen
  \bibfield  {author} {\bibinfo {author} {\bibfnamefont {A.~R.}\ \bibnamefont
  {Calderbank}}\ and\ \bibinfo {author} {\bibfnamefont {P.~W.}\ \bibnamefont
  {Shor}},\ }\href@noop {} {\bibfield  {journal} {\bibinfo  {journal} {Phys.
  Rev. A}\ }\textbf {\bibinfo {volume} {54}},\ \bibinfo {pages} {1098}
  (\bibinfo {year} {1996})}\BibitemShut {NoStop}%
\bibitem [{\citenamefont {Steane}(1996)}]{Steane:1996}%
  \BibitemOpen
  \bibfield  {author} {\bibinfo {author} {\bibfnamefont {A.~M.}\ \bibnamefont
  {Steane}},\ }\href@noop {} {\bibfield  {journal} {\bibinfo  {journal} {Phys.
  Rev. Lett.}\ }\textbf {\bibinfo {volume} {77}},\ \bibinfo {pages} {793}
  (\bibinfo {year} {1996})}\BibitemShut {NoStop}%
\end{thebibliography}

%

\newpage
\onecolumngrid

\appendix*

\section{Derivation of Equation (\ref{longeq})}\label{derivation}
Here we give a step-by-step derivation of Equation (\ref{longeq}).
The global phase factor $e^{i\theta}$ is omitted in the following equations.

\begin{align*}
&Q^{\dag}X^{\boldsymbol{e}_X}Z^{\boldsymbol{e}_Z}Q\ket{0}^{\otimes 2(n-k)}_X\ket{\psi}\\
&= Q^{\dag}X^{\boldsymbol{e}_X}Z^{\boldsymbol{e}_Z}\left(\sum_{\mu \in \mathbb{F}_2^{2(n-k)}}\ket{\mu A}\bra{\mu A}
\otimes X^{\mu N_X}Z^{\mu N_Z}\right)\ket{0}^{\otimes 2(n-k)}_X\ket{\psi}\notag\\
&= 2^{k-n}
Q^{\dag}X^{\boldsymbol{e}_X}Z^{\boldsymbol{e}_Z}\sum_{\mu \in \mathbb{F}_2^{2(n-k)}}\ket{\mu A}\otimes X^{\mu N_X}Z^{\mu N_Z}\ket{\psi}\\
&= 2^{k-n}Q^{\dag}\sum_{\mu \in \mathbb{F}_2^{2(n-k)}}X^{{\boldsymbol{e}_X}_l}Z^{{\boldsymbol{e}_Z}_l}\ket{\mu A}
\otimes (-1)^{\mu N_Z{{\boldsymbol{e}_X}_r}^T + \mu N_X{{\boldsymbol{e}_Z}_r}^T}
X^{\mu N_X}Z^{\mu N_Z}X^{{\boldsymbol{e}_X}_r}Z^{{\boldsymbol{e}_Z}_r}\ket{\psi}\\
&= 2^{k-n}\left(\sum_{\lambda \in \mathbb{F}_2^{2(n-k)}}\ket{\lambda A}\bra{\lambda A}
\otimes (X^{\lambda N_X}Z^{\lambda N_Z})^{\dag}\right)
\sum_{\mu \in \mathbb{F}_2^{2(n-k)}}X^{{\boldsymbol{e}_X}_l}Z^{{\boldsymbol{e}_Z}_l}\ket{\mu A}\\
&\quad\otimes (-1)^{\mu N_Z{{\boldsymbol{e}_X}_r}^T + \mu N_X{{\boldsymbol{e}_Z}_r}^T}
X^{\mu N_X}Z^{\mu N_Z}X^{{\boldsymbol{e}_X}_r}Z^{{\boldsymbol{e}_Z}_r}\ket{\psi}\\
&= 2^{k-n}\sum_{\mu \in \mathbb{F}_2^{2(n-k)}}(-1)^{\mu N_Z{{\boldsymbol{e}_X}_r}^T+\mu N_X{{\boldsymbol{e}_Z}_r}^T}
X^{{\boldsymbol{e}_X}_l}Z^{{\boldsymbol{e}_Z}_l}\ket{\mu A}
\otimes \left(X^{\left(\mu+{\boldsymbol{e}_X}_l A^{-1}\right)N_X}Z^{\left(\mu+{\boldsymbol{e}_X}_l A^{-1}\right)N_Z}\right)^{\dag}
X^{\mu N_X}Z^{\mu N_Z}X^{{\boldsymbol{e}_X}_r}Z^{{\boldsymbol{e}_Z}_r}\ket{\psi}\\
&= 2^{k-n}\sum_{\mu \in \mathbb{F}_2^{2(n-k)}}(-1)^{\mu N_Z{{\boldsymbol{e}_X}_r}^T+\mu N_X{{\boldsymbol{e}_Z}_r}^T}
X^{{\boldsymbol{e}_X}_l}Z^{{\boldsymbol{e}_Z}_l}\ket{\mu A}
\otimes Z^{\mu N_Z+{\boldsymbol{e}_X}_l A^{-1}N_Z}X^{\mu N_X+{\boldsymbol{e}_X}_l A^{-1}N_X}
X^{\mu N_X}Z^{\mu N_Z}X^{{\boldsymbol{e}_X}_r}Z^{{\boldsymbol{e}_Z}_r}\ket{\psi}\notag\\
&=  2^{k-n}\sum_{\mu \in \mathbb{F}_2^{2(n-k)}}(-1)^{\mu N_Z{{\boldsymbol{e}_X}_r}^T+\mu N_X{{\boldsymbol{e}_Z}_r}^T}
X^{{\boldsymbol{e}_X}_l}Z^{{\boldsymbol{e}_Z}_l}\ket{\mu A}
\otimes Z^{\mu N_Z+{\boldsymbol{e}_X}_l A^{-1}N_Z}X^{{\boldsymbol{e}_X}_l A^{-1}N_X}
Z^{\mu N_Z}X^{{\boldsymbol{e}_X}_r}Z^{{\boldsymbol{e}_Z}_r}\ket{\psi}\\
&= 2^{k-n}
\sum_{\mu \in \mathbb{F}_2^{2(n-k)}}(-1)^{\mu N_Z{{\boldsymbol{e}_X}_r}^T+\mu N_X{{\boldsymbol{e}_Z}_r}^T+\mu N_Z \left({\boldsymbol{e}_X}_l A^{-1}N_X\right)^T}
X^{{\boldsymbol{e}_X}_l}Z^{{\boldsymbol{e}_Z}_l}\ket{\mu A}\\
&\quad\otimes Z^{\mu N_Z+{\boldsymbol{e}_X}_l A^{-1}N_Z}
Z^{\mu N_Z}X^{{\boldsymbol{e}_X}_l A^{-1}N_X}X^{{\boldsymbol{e}_X}_r}Z^{{\boldsymbol{e}_Z}_r}\ket{\psi}\\
&= 2^{k-n}
\sum_{\mu \in \mathbb{F}_2^{2(n-k)}}(-1)^{\mu N_Z{{\boldsymbol{e}_X}_r}^T+\mu N_X{{\boldsymbol{e}_Z}_r}^T+\mu N_Z \left({\boldsymbol{e}_X}_l A^{-1}N_X\right)^T}
X^{{\boldsymbol{e}_X}_l}Z^{{\boldsymbol{e}_Z}_l}\ket{\mu A}
\otimes Z^{{\boldsymbol{e}_X}_l A^{-1}N_Z}
X^{{\boldsymbol{e}_X}_l A^{-1}N_X}X^{{\boldsymbol{e}_X}_r}Z^{{\boldsymbol{e}_Z}_r}\ket{\psi}\\
&= 2^{k-n}
\sum_{\mu \in \mathbb{F}_2^{2(n-k)}}
(-1)^{\mu N_Z{{\boldsymbol{e}_X}_r}^T+\mu N_X{{\boldsymbol{e}_Z}_r}^T+\mu N_Z \left({\boldsymbol{e}_X}_l A^{-1}N_X\right)^T+\mu A{{\boldsymbol{e}_Z}_l}^T}
X^{{\boldsymbol{e}_X}_l}\ket{\mu A}
\otimes Z^{{\boldsymbol{e}_X}_l A^{-1}N_Z}
X^{{\boldsymbol{e}_X}_l A^{-1}N_X}X^{{\boldsymbol{e}_X}_r}Z^{{\boldsymbol{e}_Z}_r}\ket{\psi}\\
&= X^{{\boldsymbol{e}_X}_l}\ket{{{\boldsymbol{e}_Z}_l}^T + A^{-1}N_Z{{\boldsymbol{e}_X}_r}^T + A^{-1}N_X{{\boldsymbol{e}_Z}_r}^T
+A^{-1}N_Z({\boldsymbol{e}_X}_l A^{-1}N_X)^T}_X
Z^{{\boldsymbol{e}_X}_l A^{-1}N_Z}
X^{{\boldsymbol{e}_X}_l A^{-1}N_X}X^{{\boldsymbol{e}_X}_r}Z^{{\boldsymbol{e}_Z}_r}\ket{\psi}\\
&= \ket{A^{-1}\left(A{{\boldsymbol{e}_Z}_l}^T + N_Z{{\boldsymbol{e}_X}_r}^T + N_X{{\boldsymbol{e}_Z}_r}^T + N_Z{N_X}^T{\left({\boldsymbol{e}_X}_l A^{-1}\right)}^T\right)}_X
\otimes X^{{\boldsymbol{e}_X}_l A^{-1}N_X + {\boldsymbol{e}_X}_r}Z^{{\boldsymbol{e}_X}_l A^{-1}N_Z + {\boldsymbol{e}_Z}_r}\ket{\psi}.
\end{align*}

\end{document}